\documentclass[11pt,twoside]{article}  
\usepackage{adassconf}

\begin{document}   

%
%
%

\paperID{P.122}

%
%
%
%

\title{Bibliographic Classification using the ADS Databases}

%
%
%

\author{Alberto Accomazzi,
Michael J. Kurtz,
G\"unther Eichhorn,
Edwin Henneken,
Carolyn S. Grant,
Markus Demleitner,
Stephen S. Murray}
\affil{Harvard-Smithsonian Center for Astrophysics, 60 Garden Street, Cambridge, MA 02138}

%
%

\contact{Alberto Accomazzi }
\email{aaccomazzi@cfa.harvard.edu }

%
%
%
%
%

\paindex{Accomazzi, A. }
\aindex{Kurtz, M. J.}
\aindex{Eichhorn, G.}
\aindex{Henneken, E.}
\aindex{Grant, C. S.}
\aindex{Demleitner, M.}
\aindex{Murray, S. S.}

%
%

\authormark{Accomazzi et al.}

%
%



\begin{abstract}          

We discuss two techniques used to characterize bibliographic records based
on their similarity to and relationship with the contents of the
NASA Astrophysics Data System (ADS) databases.
The first method has been used to classify input text as
being relevant to one or more subject areas based on an analysis of
the frequency distribution of its individual words.
The second method has been used to classify existing records as
being relevant to one or more databases based on the distribution
of the papers citing them. Both techniques have proven to be valuable
tools in assigning new and existing bibliographic records to different
disciplines within the ADS databases.

\end{abstract}

%
%

\section{Overview}

The NASA Astrophysics Data System (ADS; Kurtz et al 2000) 
maintains three main databases
of scientific bibliographies: Astronomy, Physics,
and the ArXiv e-prints.
Over the past few years the ADS has created and maintained a separate
``general'' database containing records which do not readily fit in
the three main databases.  The use for the general database is
twofold: it servers as a staging area for bibliographic records which may
be later incorporated into one of the other databases and it provides
a placeholder for those records which, while not being directly related
to physics or astronomy, may be cited by or citing them.  For instance,
it is not unusual for physics papers to cite articles in chemistry
or computer science and vice versa.  The typical
use of such a database is to store all records from inter-disciplinary
journals such as {\it Science} and {\it Nature}.  While some of the
articles published in these journals will be entered in the Astronomy
and Physics databases, their full table of contents will always be
available in the general database.

When new records are provided to the ADS without any meta data enabling
them to be reliably labeled as belonging to either physics or astronomy
or physics, a decision has to be made in terms of which
database they should be assigned to.  Given the sheer amount of
bibliographic data being handled by the ADS project
(Grant et al 2000), this decision
has to be made automatically most of the time.  This paper describes how
we have made use of two different tools to help us with the automatic
classification of bibliographic records.  The first tool is a
text classifier which performs an analysis of textual data based
on a well-known Bayesian probabilistic model (McCollum and Nigam 1998).
Classification of a document is performed by estimating
the likelihood of its membership in a certain database based on the
relative frequency of the words from the text in that database.
The second tool is a citation classifier which assigns existing ADS
records to one or more databases based on how frequently they have
been cited by the records in those databases.
The underlying assumption of the  citation classifier is that any papers
which have been frequently cited by papers in a particular subject area
should be considered relevant to such subject area.

Both classifiers have been trained on a set of 400 articles
published in the journal Nature during 1987.  In this sample, 39 records
were picked as being relevant to astronomy by a librarian.  The classifiers
were tested against the full set of articles published by Nature in 1997
(4033 records, 434 of which had citation data).  These records consist
of scientific research articles as well as short news, editorials,
book reviews, and obituaries.

\section{ The Text Classifier}

The problem of text classification can be summarized as follows:
given a certain string of words from a document, which of a finite
set of categories can this document be best assigned to?
Following a probabilistic approach, we chose to implement a
Multinomial Naive Bayesian Classifier
which allows a straightforward
computation of the category with the maximum likelihood based on 
the frequencies of the document's words within each category.
In our application, each category represents the set of documents
in a particular database.  Since the frequencies of the words in
each database are readily available from the database-specific indexes
that the ADS maintains, the computation of the probabilities
can be carried out in real time from the index data.

The implementation of the classifier
showed that it performed well in classifying documents for which
at least 20 text words were available from either the title or the
abstract.
The challenge we were
presented was trying to classify records for which only a title was
available.  In order to improve the classifier, a number of pre- and
post-scoring steps were taken:

\begin{itemize}
\item The input text was pre-processed using the standard parsing
rules used by the ADS search engine (Accomazzi et al 2000).

\item All words consisting solely of digits were removed, as well
as title words and phrases which had no relevance for classification
(e.g. ``obituary'').

\item The likelihood score generated by the Bayesian classifier was
normalized in order to limit the contribution of records consisting of
few words, for which the highest rate of misclassification was found.

\item To compensate for the previous step, a set of
database-specific ``trigger'' keywords were defined which, when found,
boosted the classification score of the input text.
\end{itemize}

The resulting classifier was implemented as a two-parameter function:
$N_t$, the minimum number of words required for
a document to be considered classifiable, and $S_t$, the minimum
classification score necessary for a document to be considered belonging
to a particular database.
The results of the classification are displayed in Figure~\ref{fig1}, where
the performance of the classifier can be judged by looking at
the Precision ($P$) and Recall ($R$) of the classification for each input set
of cutoff scores and minimum number of words.  
As one can see from the plot, the classifier has been designed to yield a
high precision irregardless of the number of input words.  This is a crucial
issue for our application since we do not want to mistakenly assign
non-relevant records to any of the ADS databases.

\section{The Citation Classifier}

The citation classifier was implemented to assign existing ADS
records to one or more databases based on how frequently they have
been cited by the records in those databases.
The underlying assumption of the citation classifier is that any papers
which have been frequently cited by papers in a particular subject area
should be considered relevant to such subject area (Kurtz et al 2002).
The scope and usefulness of this classifier is obviously limited by
the availability of citation data for the records being considered:
an article which has not been cited in any of the astronomy or
physics journals for which the ADS has reference data cannot be
categorized by the classifier.  However, since the coverage of
journal references from the core astronomy literature is
quite thorough in the ADS, we can expect that important research
articles will be cited with some frequency within astronomy.

Based on this premise, we implemented a citation classifier
by considering, for each record for which citations are available,
the ratio between the number of
citations belonging to a particular database and the total number of citations.
If the ratio is high enough we can conclude that since a significant
portion of the papers citing the record in question come from a single
database, the paper in question is relevant to that database.
The citation classifier was implemented as a function taking as
input two parameters: $N_c$, the minimum number of citations required
for a record to be considered classifiable, and $R_c$, the ratio
between the number of citations in a particular database and the total
number of citations.

\begin{figure}
\plotone{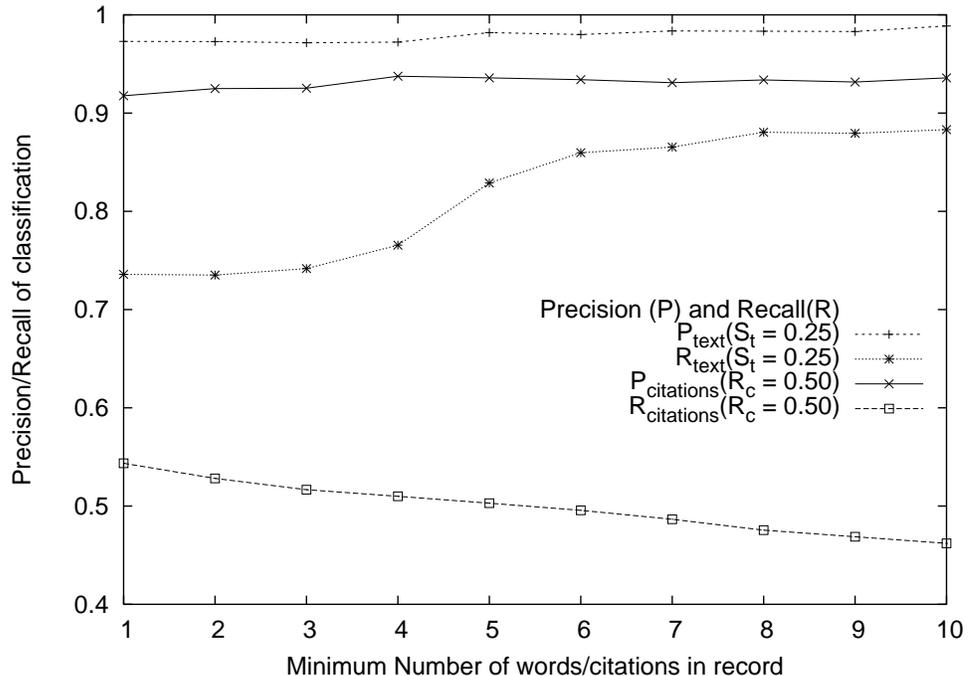}
\caption{Results for the text and citation classifiers} \label{fig1}
\end{figure}

The performance of the citation classifier was tested against
the set of 434 articles in the Nature sample which
have citation data available.  The results of the classifier are
summarized in Figure~\ref{fig1}.
Once again we notice little variation in the performance of the classifier
as a function of the total number of citations for a given paper, which
is a desirable feature.  On the other hand, given the limited number
of citations for some of the records available, the recall is
much lower than what was achieved with the text classifier.

\section{Discussion}

The text and citation classifiers described here have shown to
be a valuable tool in categorizing records from scientific journals
such as Nature and Science for the purpose of introducing them into
the Astronomy or Physics databases.
Further inspection of the results showed that 
misclassified papers are often borderline cases involving subjects
such as Geophysics and Planetary Science which overlap the different
databases.  Additionally, a small number of records which were originally not
selected as belonging to Astronomy by the librarian were later
found to be relevant upon a subsequent review of the results by
the classifiers.

Because the text and citation classifiers use
different data when assigning articles to a database, we find that
best overall results can be achieved by combining
the output from both classifiers.  By choosing
conservative settings for the parameters controlling the classifiers
($S_t = 0.25, M_t = 5, R_c = 0.5, N_c = 4$).
we were able to achieve a precision of 0.94 with a
recall of 0.89 when classifying the sample against
the astronomy database.

\acknowledgments
The ADS is funded by NASA Grant NCC5-189 and is available online at
\htmladdURL{http://ads.harvard.edu}

\end{document}